\documentclass{desyproc}

\begin{document}

\title{Axion cooling of white dwarfs}

\author{{\slshape J. Isern$^{1,2}$, 
                  S. Catal\'an$^3$,
                  E. Garc\'{\i}a--Berro$^{4,2}$, 
                  M. Salaris$^5$, 
                  S. Torres$^{4,2}$ }\\[1ex]
          $^1$Institut de Ci\`encies de l'Espai (CSIC), 
              Campus UAB, 08193 Bellaterra, Spain \\
          $^2$Institut d'Estudis Espacials de Catalunya (IEEC), 
              Ed. Nexus, c/Gran Capit\`a, 08034 Barcelona, Spain\\
          $^3$Center for Astrophysics Research, University of Hertfordshire, 
              College Lane, Hatfield AL10 9AB, UK \\
          $^4$Departament de F\'{\i}sica Aplicada, 
              Universitat Polit\`ecnica de Catalunya,
              c/Esteve Terrades 5, 08860 Castelldefels, Spain\\
          $^5$Astrophysics Research Institute, 
              Liverpool John Moores University, 
              12 Quays House, Birkenhead, CH41 1LD, UK \\ }

\contribID{Isern\_Jordi}

\desyproc{DESY-PROC-2012-04}
\acronym{Patras 2012}
\doi

\maketitle

\begin{abstract}
The evolution of white dwarfs  is a simple gravothermal process.  This
process can be tested in  two ways, through the luminosity function of
these  stars  and through  the  secular  variation  of the  period  of
pulsation of those stars that are  variable. Here we show how the mass
of  the axion  can be  constrained  using the  white dwarf  luminosity
function.
\end{abstract}

\section{Introduction}

White  dwarfs are  the final  remnants of  low-  and intermediate-mass
stars  \cite{review}.   Since  their  structure is  dominated  by  the
pressure of  partially or  strongly degenerate electrons,  they cannot
obtain  energy  from the  thermonuclear  burning  processes and  their
evolution  can be  described, in  a first  approximation, as  a simple
gravothermal process.  The white dwarf cooling rate can be measured in
two  ways, using their  luminosity function  or employing  the secular
variation of  the period  of pulsation in  the case of  variable white
dwarfs.

An important  property of the  luminosity function of white  dwarfs is
that  the  shape of  the  bright branch  is  independent  of the  star
formation rate and provides a  direct measurement of the mass averaged
characteristic  cooling  time  \cite{Isern:2008}.  This  property  has
provided some hint that axions of the DFSZ type could exist and should
have  a mass in  the range  of $\sim  4-8$ ~meV,  but not  much larger
\cite{Isern:2008a,Isern:2009,Isern:2012}.

\section{The luminosity function}

\begin{figure}[t]
\includegraphics[width=0.45\textwidth]{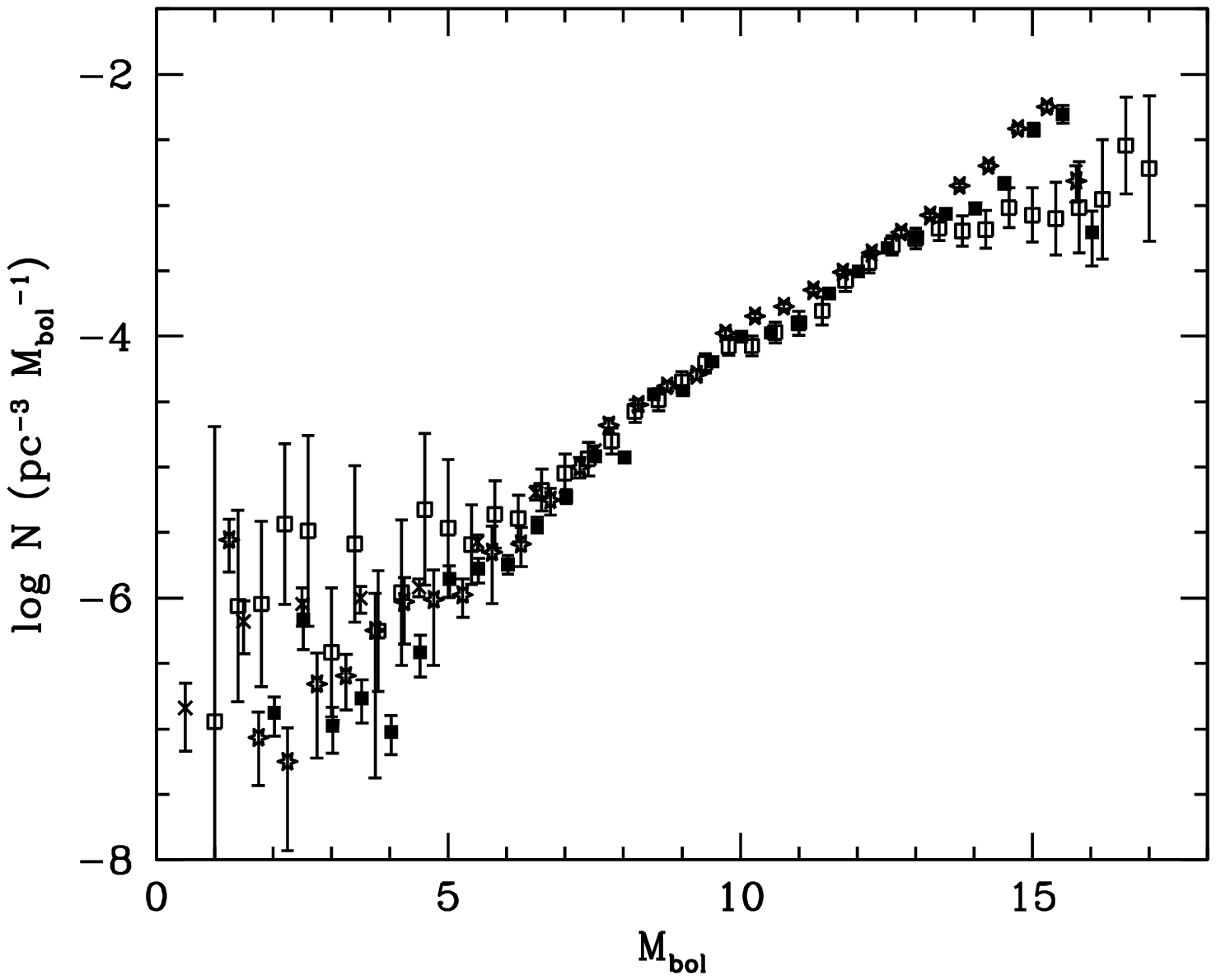}\hspace*{0.5cm}
\includegraphics[width=0.45\textwidth]{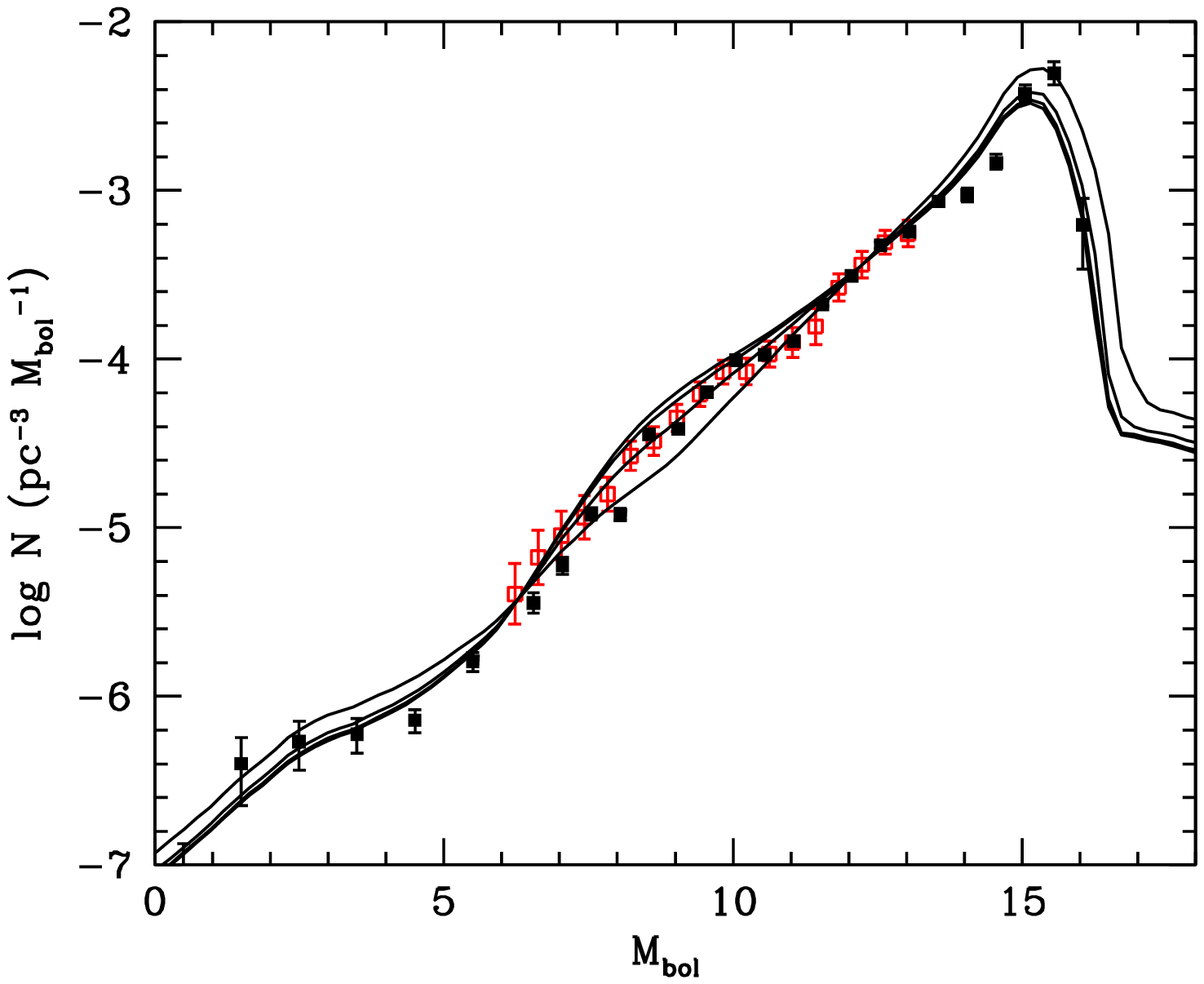}
\caption{Left: Luminosity  functions  obtained  from large  surveys.   Solid
  squares:  SDSS,   all  spectral  types   \cite{Harris:2006},  hollow
  squares: SDSS,  only DA  white dwarfs \cite{dege08},  crosses: SDSS,
  hot DA  white dwarfs \cite{Krzesinski:2009},  stars: SuperCOSMOS Sky
  Survey \cite{rowe11}. Right: Luminosity  functions  obtained  assuming an  axion  electron
  coupling  (up to down)  $g_{\rm ae}  \times 10^{13}  =0.0,\, 1.12,\,
  2.24,\,  4.48$.  Black squares:  extended  luminosity function,  red
  hollow squares: pure DA luminosity function, see text for details.}
\label{Fig:newlf}
\end{figure}

The white dwarf luminosity  function has been noticeably improved with
the  data  provided  by  large sky  surveys.   Figure  \ref{Fig:newlf}, right,
displays  the luminosity function  obtained from  the SDSS  with white
dwarfs identified from  their proper motion \cite{Harris:2006} (HA-LF)
and  the  one  constructed  with   the  same  method  but  limited  to
spectroscopically-identified    DA    white    dwarfs    \cite{dege08}
(DG-LF). The  monotonic behavior of this function  clearly proves that
the evolution of white dwarfs  is a simple gravothermal process, while
the sharp cut-off shown by  the HA-LF distribution at low luminosities
is the consequence of the  finite age of the Galaxy. The discrepancies
between  both at  low  luminosities  are well  understood  and can  be
attributed to  the different way  in which the  effective temperatures
and    gravities    of     the    sample    have    been    determined
\cite{dege08}. Furthermore,  the DG-LF only considers  DA white dwarfs
and,  at low temperatures,  it is  hard to  separate these  stars from
non-DA white dwarfs.   For this reason the DG-LF has  to be limited to
magnitudes smaller than $M_{\rm  bol} \sim 13$.  At high luminosities,
$M_{\rm  bol} <6$,  the  dispersion of  both functions  is
considerably large.  The reason is that both luminosity functions have
been  built  using the  reduced  proper  motion  method which  is  not
appropriate for bright white dwarfs. The UV-excess method has
allowed  to build a  luminosity function  for magnitudes  ranging from
$-0.75$ to 7 (KZ-LF) \cite{Krzesinski:2009}.  This method, however, is
not adequate for dim stars  and becomes rapidly incomplete out of this
range of magnitudes. Since this  sample overlaps with the HA-LF, it is
possible  to extend the  luminosity function  to the  brightest region
assuming continuity and  just retaining the KZ-LF data  in this region
(black squares  of Fig.~\ref{Fig:newlf}), right.  At  this point we  recall that
the process of  formation of white dwarfs is not  well known and that their
structure is somewhat dependent on  the initial  conditions until
 neutrino losses have finished to erase the memory of these conditions.
This happens when  $M_{\rm bol}
\sim  6$   or,  equivalently,  when  $\log   (L/L_\odot)  \sim  -0.5$.
Therefore, the  use of very  hot white dwarfs for  diagnostic purposes
has to be taken with some care.

Since the  integration time  of the SDSS  catalogue is fixed,  the S/N
ratio depends  on the brightness  of the source  and this can  lead to
large uncertainties  in the determination  of the parameters  of faint
white  dwarfs  with  the  consequent risk  of  introducing  systematic
errors. Fortunately, a  completely independent luminosity function has
been recently  obtained \cite{rowe11} from the  SuperCOSMOS Sky Survey
(hereafter called  RH-LF) that does  not excessively deviate  from the
HA-LF at low luminosities.  This means that these luminosity functions
are probably  not affected by large  systematic effects. Additionally,
it also indicates  that neither the shape of the  peak nor the cut-off
are  yet well  constrained. It is important to notice here
that in  this region the  luminosity function strongly depends  on the
star formation rate. Furthermore, since  the RH-LF  uses the  proper motion
method  to cull  white dwarfs,  the bright  branch of  this luminosity
function suffers  from the same drawbacks  as the HA-LF  and the DG-LF
luminosity functions.   

\section{Results and conclusions}

Figure~\ref{Fig:newlf}, tight,  displays   the  theoretical  luminosity  function
computed  with  the method  described  in Ref.~\cite{Isern:2008},  but
using   updated   evolutionary   sequences   for   DA   white   dwarfs
\cite{Salaris:2010}.  As can be seen, the computed luminosity function
reproduces reasonably well the region  where the slope only depends on
the characteristic cooling time except in the region $M_{\rm bol} \sim
10$, where the computed value predicts an excess of white dwarfs. This
region corresponds  to the region where neutrino  emission has already
stopped  and  axion emission,  if  included,  still  persists.  If  we
restrict ourselves to the region  where the luminosity function of DAs
is reliable  (red squares in  Fig.~\ref{Fig:newlf}), right, the best  fit occurs
for  $g_{\rm  ae}  \sim  2.2  \times  10^{-13}$  and  is  clearly  not
satisfactory for  $g_{\rm ae}  > 5 \times  10^{-13}$.  This  effect is
tiny and could  be due to many conventional  effects (like the details
of  envelope modelling, or  the choice  of the  initial to  final mass
relationship, IMF, or  metallicity, or even to a  recent burst of star
formation  \cite{Isern:2009}).  All  these  alternatives  need  to  be
systematically examined.

Recently, it has been  claimed \cite{Melendez:2012} that the inclusion
of the axion emission distorts the thermal profile of white dwarfs and
this  weakens the  ability  of the  luminosity  to bound  the mass  of
axions.   The change  of the  thermal profile  is a  well-known effect
produced by  any important emission  process, and this is  indeed what
occurs in the  case of bright white dwarfs  and/or large axion masses.
However,  these  white  dwarfs  are  not  appropriate  for  diagnostic
purposes and  here we only  consider axions with rather  small masses.
Thus, the  approach adopted  here is enough  for our purposes.  In any
case, the corresponding bounds do not differ much, and the differences
can  possibly  be  attributed  to  the  different  way  in  which  the
luminosity function is computed.

In this  sense, the  new measurement of  the secular evolution  of the
pulsation  period  of  G117-B15A  \cite{Kepler:2011} has  provided  an
additional support to the hypothesis  that white dwarfs are cooling at
a rate larger than expected,  and that axions could be the responsible
of  this  \cite{Isern:1992,Isern:2010}.  However,  a  recent  detailed
analysis  \cite{Corsico:2012a}  suggests  that the  coupling  constant
necessary to account for the  observations is $g_{\rm ae} = 4.9 \times
10^{-13}$ or equivalently  $m_{\rm a} \approx 17$ meV. Recently, it
has been found \cite{Corsico:2012b}  that  the rate of
period change of the variable white dwarf R548 (ZZ Ceti itself) yields
$g_{\rm ae}  = 4.5 \times 10^{-13}$  or $m_{\rm a} \approx  16$ meV, a
very  similar result. These values are larger than the upper bound  
obtained from the luminosity function and could also be in conflict with
other bounds like SN1987A \cite{Raffelt:2011}.   Nevertheless, in all cases, 
the uncertainties are still large enough to prevent a definite conclusion
and an  improvement of both the models and observational data is
necessary.

\section*{Acknowledgments}

This    work   has    been    supported   by    the   MICINN    grants
AYA2011-24704-1839/ESP and AYA2011-23102, by the ESF EUROCORES Program
EuroGENESIS (MICINN  grant EUI2009-04170 and 04167), by  SGR grants of
the Generalitat de Catalunya and by the European Union FEDER funds.

\begin{footnotesize}

\end{footnotesize}

\end{document}